# Quantum Machine Learning Approach for the Prediction of Surface Roughness in Additive Manufactured Specimens

Akshansh Mishra[1,2*], Vijaykumar S. Jatti[3]

[1]School of Industrial and Information Engineering, Politecnico di Milano, Milan, Italy

[2]Principal AI Scientist, Neural Net, India

[3] Department of Mechanical Engineering, Symbiosis Institute of Technology, Pune, India

**Abstract:** Surface roughness is a crucial factor influencing the performance and functionality of additive manufactured components. Accurate prediction of surface roughness is vital for optimizing manufacturing processes and ensuring the quality of the final product. Quantum computing has recently gained attention as a potential solution for tackling complex problems and creating precise predictive models. In this research paper, we conduct an in-depth comparison of three quantum algorithms – the Quantum Neural Network (QNN), Quantum Forest (Q-Forest), and Variational Quantum Classifier (VQC) adapted for regression – for predicting surface roughness in additive manufactured specimens for the first time. We assess the algorithms' performance using Mean Squared Error (MSE), Mean Absolute Error (MAE), and Explained Variance Score (EVS) as evaluation metrics. Our findings show that the Q-Forest algorithm surpasses the other algorithms, achieving an MSE of 56.905, MAE of 7.479, and an EVS of 0.2957. In contrast, the QNN algorithm displays a higher MSE of 60.840 and MAE of 7.671, coupled with a negative EVS of -0.444, indicating that it may not be appropriate for predicting surface roughness in this application. The VQC adapted for regression exhibits an MSE of 59.121, MAE of 7.597, and an EVS of -0.0106, suggesting its performance is also inferior to the Q-Forest algorithm.

1. Introduction

Quantum computing, an innovative field that draws on the principles of quantum mechanics, has attracted substantial interest in recent years due to its capacity to revolutionize various industries. Unlike classical computing, which uses bits to represent information as either 0 or 1, quantum computing employs qubits that can coexist in multiple states simultaneously, thanks to the phenomena of superposition and entanglement [1-5]. This unique characteristic allows quantum computers to perform intricate calculations at an exponentially faster rate than conventional computers, creating opportunities to solve previously unsolvable problems.

The manufacturing industry, characterized by its elaborate and complex processes, stands to benefit significantly from advancements in quantum computing. Potential applications of quantum computing in manufacturing cover a wide range of areas, including optimization, simulation, material science, and predictive modeling. Integrating quantum computing into these fields could lead to considerable improvements in efficiency, cost reduction, and product quality [6-9].

Quantum computing can optimize various manufacturing processes, such as scheduling, resource allocation, and supply chain management, by identifying the most efficient and cost-effective solutions. Quantum optimization algorithms, like the Quantum Approximate Optimization Algorithm (QAOA) and Quantum Adiabatic Computing, hold the potential to outperform classical optimization techniques in tackling complex, large-scale problems. Quantum computers can simulate quantum systems more efficiently than classical computers, allowing researchers to gain a deeper understanding of materials and chemicals at the atomic and molecular levels. This enhanced understanding could accelerate the discovery of new materials and the development of cutting-edge manufacturing techniques, resulting in more sustainable and efficient production processes.

In the realm of material science, quantum computing can expedite the discovery and design of novel materials with customized properties by simulating their behavior at the quantum level. This could lead to the creation of advanced materials with improved strength, durability, and energy efficiency, thereby enhancing the overall quality of manufactured products [10-13]. Quantum computing can also significantly improve predictive modeling in manufacturing by providing more accurate and efficient solutions for intricate problems. Quantum machine learning algorithms, such as the Quantum Neural Network (QNN), Quantum Forest (Q-Forest), and Variational Quantum Classifier (VQC), can be utilized to predict critical parameters, like surface roughness, machine wear, and defect rates, which influence product quality and process efficiency. These algorithms offer the potential to decrease manufacturing errors, minimize downtime, and optimize maintenance schedules.

In the present work, we report the first-ever implementation of Quantum Machine Learning algorithms for determining the surface roughness of additive manufactured specimens. This pioneering approach demonstrates the potential of harnessing quantum computing techniques to enhance the accuracy and efficiency of predictive modeling in the field of additive manufacturing.

## 2. Quantum Machine Learning Framework

Quantum machine learning is an emerging interdisciplinary field that integrates quantum computing with machine learning, aiming to harness the unique properties of quantum systems for more efficient complex calculations and optimization problem-solving than classical computing. As this field is still in its nascent stage, its potential applications are vast and varied. To comprehend the intricate workings of quantum machine learning, it is essential to understand several fundamental concepts of Quantum bits (qubits), Quantum entanglement, and Quantum gates.

In quantum computing, qubits are the basic units that can exist in a superposition of both 0 and 1 states, which is unlike classical bits that can only exist in either 0 or 1 state. Qubits are represented as complex linear combinations of the basis states $|0\rangle$ and $|1\rangle$ shown in Figure 1. This property enables parallelism and allows quantum computers to process extensive information concurrently.

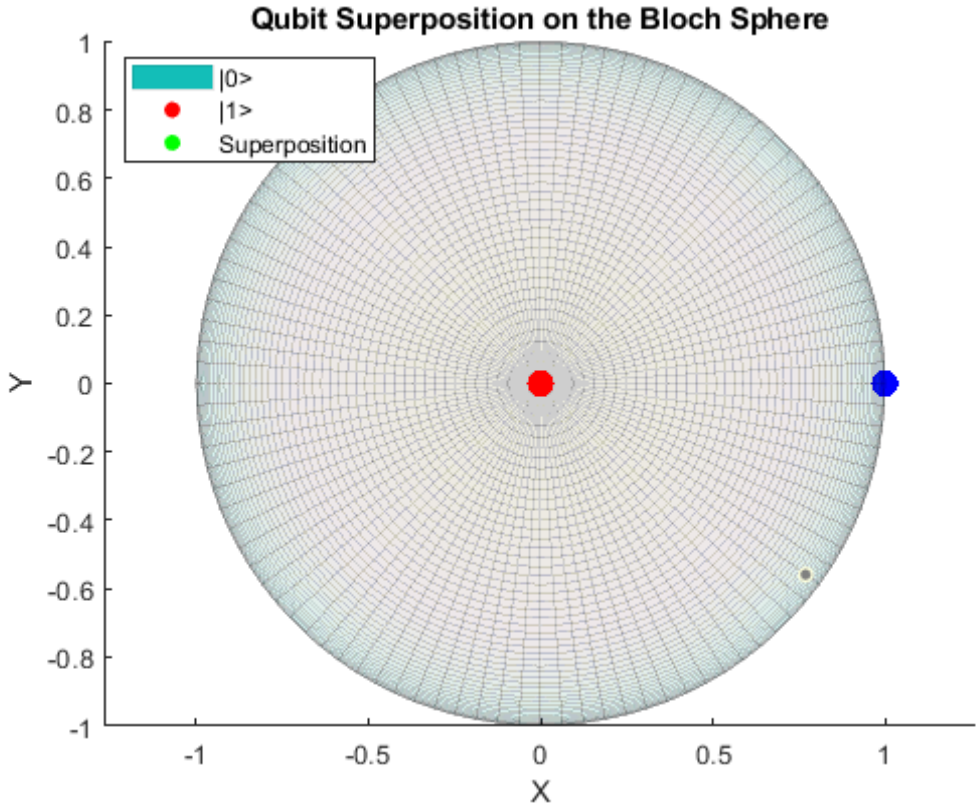

Figure 1: Schematic representation of Quantum Superposition

To explain this with equations, let's consider a qubit, represented as |ψ⟩ as depicted in Equation 1.

$$|\psi\rangle = \alpha|0\rangle + \beta|1\rangle \tag{1}$$

Here, α and β are complex coefficients that determine the probability amplitudes of the qubit being in state |0⟩ or |1⟩, respectively. The probabilities of the qubit being in state |0⟩ or |1⟩ are given by Equation 2 and 3.

$$P(|0\rangle) = |\alpha|^2 \tag{2}$$

$$P(|1\rangle) = |\beta|^2 \tag{3}$$

Since the qubit must be in either state |0⟩ or |1⟩, the sum of the probabilities must be equal to 1 as depicted in Equation 4.

$$|\alpha|^2 + |\beta|^2 = 1 \tag{4}$$

This property of qubits allows them to exist in a superposition of states and enables quantum parallelism. In a quantum computer with n qubits, there can be $2^n$ different states that can be processed simultaneously. This is because the state of an n-qubit system can be written as Equation 5.

$$|\psi_n\rangle = \Sigma_i \, c_i \, |i\rangle \qquad (5)$$

where i ranges from 0 to $(2^n - 1)$ and $\Sigma_i \, |c_i|^2 = 1$.

This parallelism allows quantum computers to perform complex computations that are infeasible for classical computers.

Quantum entanglement is an exceptional phenomenon that occurs when two or more qubits become intricately interconnected, rendering them inseparable shown in Figure 2. This can be represented by the entangled state Equation 6.

$$|\Psi\rangle = \alpha|00\rangle + \beta|11\rangle \qquad (6)$$

where $|\Psi\rangle$ is the entangled state, α and β are complex coefficients, and $|00\rangle$ and $|11\rangle$ are the basis states of the two entangled qubits. In this scenario, the state of one qubit cannot be described independently of the other, as the probabilities of the outcomes are determined by the combined state $|\Psi\rangle$.

Entanglement is a crucial resource in quantum computing, as it enables non-local correlations and augments computational capabilities. One example of this is the Bell states, which are a set of maximally entangled states that can be written as Equation 7.

$$|\Phi+\rangle = (1/\sqrt{2})(|00\rangle + |11\rangle)$$
$$|\Phi-\rangle = (1/\sqrt{2})(|00\rangle - |11\rangle)$$
$$|\Psi+\rangle = (1/\sqrt{2})(|01\rangle + |10\rangle)$$
$$|\Psi-\rangle = (1/\sqrt{2})(|01\rangle - |10\rangle) \qquad (7)$$

These Bell states demonstrate the strong correlations that can be achieved through quantum entanglement, which can be leveraged in various quantum algorithms, communication protocols, and teleportation, thus significantly enhancing the computational power of quantum systems.

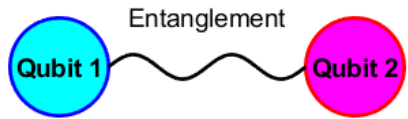

Figure 2: The entanglement between the qubits

As the foundation of quantum circuits, quantum gates perform operations on qubits. Analogous to classical logic gates (e.g., AND, OR, NOT), quantum gates are reversible and can act on quantum state superpositions shown in Figure 3.

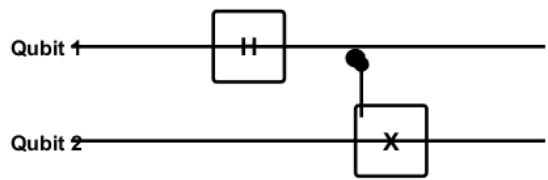

Figure 3: Quantum circuit with two qubits, a Hadamard gate (H) acting on Qubit 1, and a CNOT gate acting on both qubits

The detailed mechanism of quantum machine learning is shown in Figure 4.

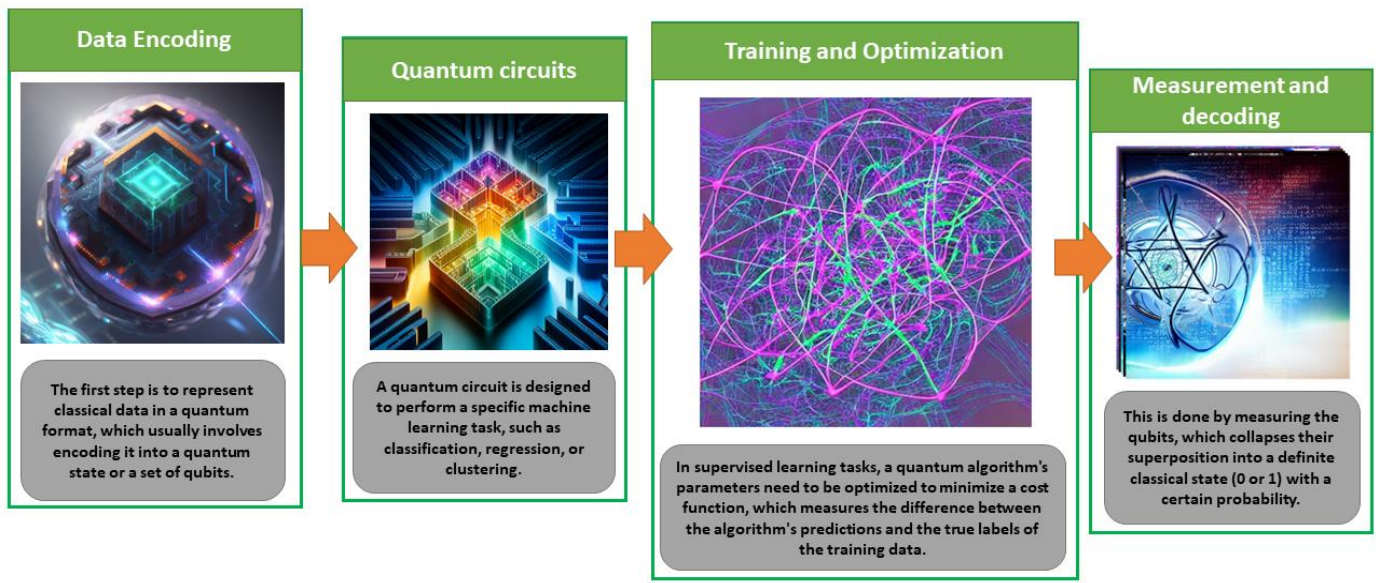

Figure 4: Framework of Quantum Machine Learning

The initial step involves converting classical data into a quantum format, typically by encoding it into a quantum state or a set of qubits. This process employs quantum feature maps or embedding techniques that transform classical data into a high-dimensional Hilbert space, making it compatible with quantum computing. A quantum circuit tailored to a specific machine learning task, such as classification, regression, or clustering, is designed by selecting suitable quantum gates and arranging them sequentially to process input data and produce the desired output. The circuit's structure and complexity may vary according to the problem and required accuracy level. For supervised learning tasks, quantum algorithm parameters must be optimized to minimize a cost function that measures the discrepancy between algorithm predictions and the true labels of the training data. Quantum algorithms often employ variational techniques in which a classical optimization algorithm updates the quantum circuit parameters based on feedback from the quantum computer. Upon processing input data through the quantum circuit, results must be extracted by measuring qubits, causing their superposition to collapse into a definite classical state (0 or 1) with specific probabilities. The classical output is then decoded and interpreted to provide insights or predictions concerning the original problem.

## 3. Materials and Methods

In order to maintain consistency in the model while reducing print size, material usage, and time, ASTM E8 standard geometry was adopted with a uniform 50% reduction in dimensions. The Response Surface Methodology (RSM) Design of Experiment was utilized to create 30 distinct trial conditions, each with three input parameter levels (see Figure 5).

Using Ultimaker Cura software, the CAD model was sliced to generate G-code (see Figure 6). The experimental investigation was carried out using the Creality 3D FDM printer (see Figure 7), with each print assigned a unique set of settings that varied in layer height, infill density, infill pattern, bed temperature, and nozzle temperature, and using Polylactic Acid (PLA) material. An input parameter datasheet was created, and the differences in length between each model and the original CAD file were measured using a digital vernier caliper.

The data obtained from the experiment has been converted into a CSV file and then imported into a Google Colab platform. This platform will be used to implement three different Quantum Machine Learning algorithms - Quantum Neural Network (QNN), Q-Forest, and Variational Quantum Classifier (VQC) - all of which have been adapted for regression analysis. The performance of these algorithms were evaluated on the basis of three metric features i.e. Mean Square Error (MSE), Mean Absolute Error (MAE) and Explained Variance Score.

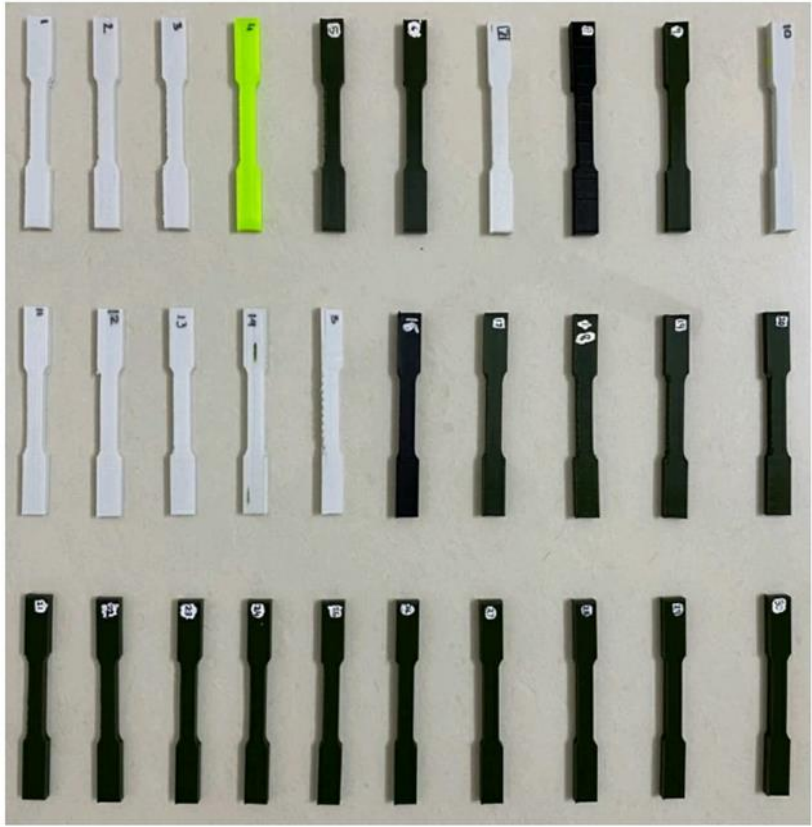

Figure 5: Fabricated Additive Manufactured specimens

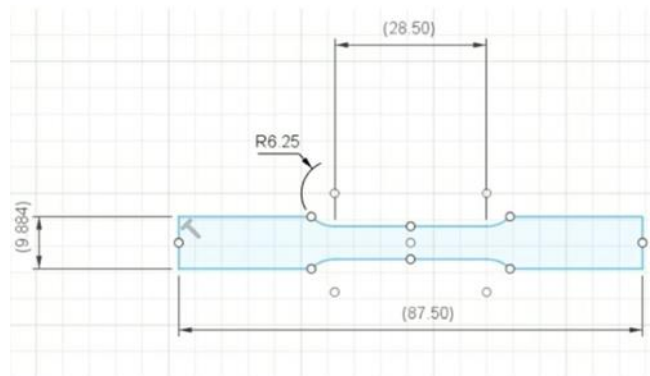

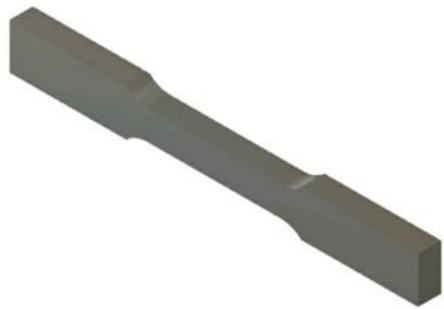

Figure 6: Schematic sketch of the specimen

## 4. Results and Discussion

The results of the Surface roughness measurements obtained through various input parameter combinations are presented in Table 1.

Table 1: Experimental results

| Layer Height (mm) | Wall Thickness (mm) | Infill Density (%) | Infill Pattern | Nozzle Temperature (°C) | Bed Temperature (°C) | Print Speed (mm/sec) | Fan Speed (%) | Surface Roughness ($\mu m$) |
|---|---|---|---|---|---|---|---|---|
| 0.1 | 1 | 50 | honeycomb | 200 | 60 | 120 | 0 | 6.12275 |
| 0.1 | 4 | 40 | grid | 205 | 65 | 120 | 25 | 6.35675 |
| 0.1 | 3 | 50 | honeycomb | 210 | 70 | 120 | 50 | 5.957 |
| 0.1 | 4 | 90 | grid | 215 | 75 | 120 | 75 | 5.92025 |
| 0.1 | 1 | 30 | honeycomb | 220 | 80 | 120 | 100 | 6.08775 |
| 0.15 | 3 | 80 | honeycomb | 200 | 60 | 60 | 0 | 6.0684 |
| 0.15 | 4 | 50 | grid | 205 | 65 | 60 | 25 | 9.27525 |
| 0.15 | 10 | 30 | honeycomb | 210 | 70 | 60 | 50 | 7.479 |
| 0.15 | 6 | 40 | grid | 215 | 75 | 60 | 75 | 7.557 |
| 0.15 | 1 | 10 | honeycomb | 220 | 80 | 60 | 100 | 8.48675 |
| 0.2 | 5 | 60 | honeycomb | 200 | 60 | 40 | 0 | 8.4695 |
| 0.2 | 4 | 20 | grid | 205 | 65 | 40 | 25 | 8.8785 |
| 0.2 | 5 | 60 | honeycomb | 210 | 70 | 40 | 50 | 9.415 |
| 0.2 | 7 | 40 | grid | 215 | 75 | 40 | 75 | 9.71375 |
| 0.2 | 3 | 60 | honeycomb | 220 | 80 | 40 | 100 | 10.59625 |
| 0.1 | 1 | 50 | triangl | 200 | 60 | 120 | 0 | 6.04925 |

| | | | es | | | | | |
|---|---|---|---|---|---|---|---|---|
| 0.1 | 4 | 40 | cubic | 205 | 65 | 120 | 25 | 9.262 |
| 0.1 | 3 | 50 | triangles | 210 | 70 | 120 | 50 | 6.127 |
| 0.1 | 4 | 90 | cubic | 215 | 75 | 120 | 75 | 5.99675 |
| 0.1 | 1 | 30 | triangles | 220 | 80 | 120 | 100 | 6.1485 |
| 0.15 | 3 | 80 | triangles | 200 | 60 | 60 | 0 | 8.2585 |
| 0.15 | 4 | 50 | cubic | 205 | 65 | 60 | 25 | 8.347 |
| 0.15 | 10 | 30 | triangles | 210 | 70 | 60 | 50 | 8.2385 |
| 0.15 | 6 | 40 | cubic | 215 | 75 | 60 | 75 | 8.23125 |
| 0.15 | 1 | 10 | triangles | 220 | 80 | 60 | 100 | 8.35125 |
| 0.2 | 5 | 60 | triangles | 200 | 60 | 40 | 0 | 9.072 |
| 0.2 | 4 | 20 | cubic | 205 | 65 | 40 | 25 | 9.23825 |
| 0.2 | 5 | 60 | triangles | 210 | 70 | 40 | 50 | 9.18225 |
| 0.2 | 7 | 40 | cubic | 215 | 75 | 40 | 75 | 9.299 |
| 0.2 | 3 | 60 | triangles | 220 | 80 | 40 | 100 | 9.382 |

Now let's discuss about the concept results obtained by the implemented algorithms individually in the below sub-sections.

*4.1 Quantum Neural Network (QNN)*

Quantum Neural Networks (QNNs) represent an innovative approach that merges the principles of quantum computing with classical neural network architectures to address complex problems more effectively. Utilizing quantum bits (qubits) as the primary information carriers, QNNs leverage the inherent quantum properties of superposition to allow for simultaneous representation of multiple states.

At the core of QNNs lie quantum layers, which consist of sequential quantum gates. These gates are the fundamental mathematical components of quantum computing, responsible for transforming qubit states. Quantum gates are represented by unitary matrices, ensuring that the normalization of qubit states is maintained. Consequently, a quantum circuit is formed through a series of quantum gates, with the overall transformation resulting from the product of the matrices that represent the individual gates.

In mathematical terms, a quantum state can be expressed as a complex vector in a Hilbert space, where each element corresponds to the probability amplitude of a distinct computational basis state. The quantum state comprising n qubits can be denoted as Equation 8.

$$|\psi\rangle = \Sigma (a_i |i\rangle) \tag{8}$$

Here, i ranges from 0 to $2^n - 1$, $a_i$ signifies complex coefficients, and $|i\rangle$ represents the corresponding computational basis state. The probability of measuring the state $|i\rangle$ is given by the square of the modulus of $a_i$.

The transformation of a quantum state through a quantum gate can be depicted as a matrix-vector multiplication as depicted in Equation 9.

$$|\psi'\rangle = U |\psi\rangle \tag{9}$$

In this equation, U represents the unitary matrix corresponding to the quantum gate, $|\psi\rangle$ denotes the initial state, and $|\psi'\rangle$ symbolizes the final state after applying the gate.

To determine the output of a QNN, the quantum circuit is executed on either a quantum computer or a simulator. In our present work, the quantum circuit is executed on a simulator as shown in Figure 7. The resulting output is a probability distribution over the computational basis states, which is subsequently processed by a classical neural network or other classical machine learning models to generate the final output.

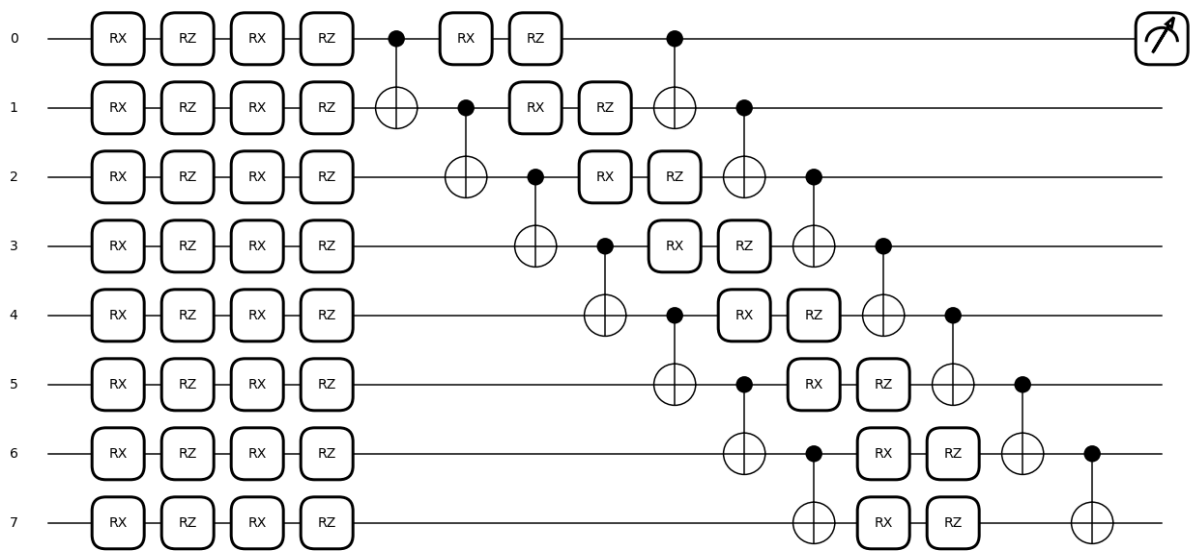

Figure 7: Quantum Circuit framework in the present work

The QNN is defined by the function **qnn(params, x=None)**, which takes two arguments: **params** representing the network's parameters, and an optional input feature vector **x**. Within the QNN, the input features are embedded into the quantum state using a series of rotations (**qml.RX** and **qml.RZ**). These rotations act on the qubits of the quantum circuit, effectively encoding the classical data into the quantum system.

The quantum layers are defined in a loop that iterates through the **params** array. For each layer, rotations are applied to the qubits using the parameters in the array, followed by a series of controlled NOT (CNOT) gates that create entanglement between adjacent qubits. This entanglement allows for the exploration of a larger solution space and enhances the network's expressivity. The output of the QNN is obtained by measuring the expectation value of the Pauli-Z operator on the first qubit. This expectation value represents the network's prediction for the given input.

The cost function, **cost(params, X, y)**, calculates the MSE between the QNN's predictions and the target values. It does this by iterating through the input data **X** and calling the QNN function with the current parameters and input features. The mean squared error between the predictions and the ground truth targets **y** is then computed and returned as the cost value.

The QNN's parameters are initialized randomly and optimized using the gradient descent optimizer provided by the Pennylane library. The optimization is performed for 100 iterations, with the parameters updated at each step to minimize the cost function. Finally, the trained QNN is used to make predictions on the test set, and the performance is evaluated using three metrics: mean squared error (MSE), mean absolute error (MAE), and the Explained Variance Score as shown in Table 2.

Table 2: Obtained Metrics features for QNN Algorithm

| MSE | MAE | Explained Variance Score |
| --- | --- | --- |
| 60.840 | 7.671 | -0.444 |

The plot of explained variance score vs training iterations is typically used to track the performance of a machine learning algorithm during the training process as shown in Figure 8. The explained variance score is a metric that measures the proportion of the variance in the target variable that is explained by the model. It ranges from 0 to 1, where 0 indicates that the model does not explain any variance, and 1 indicates that the model explains all of the variance. The plot of explained variance score vs training iterations can provide insights into how well the model is learning and converging to a solution. If the explained variance score is increasing with each iteration, it suggests that the model is improving and learning from the data. On the other hand, if the explained variance score is plateauing or decreasing, it suggests that the model is not improving or may be overfitting the data.

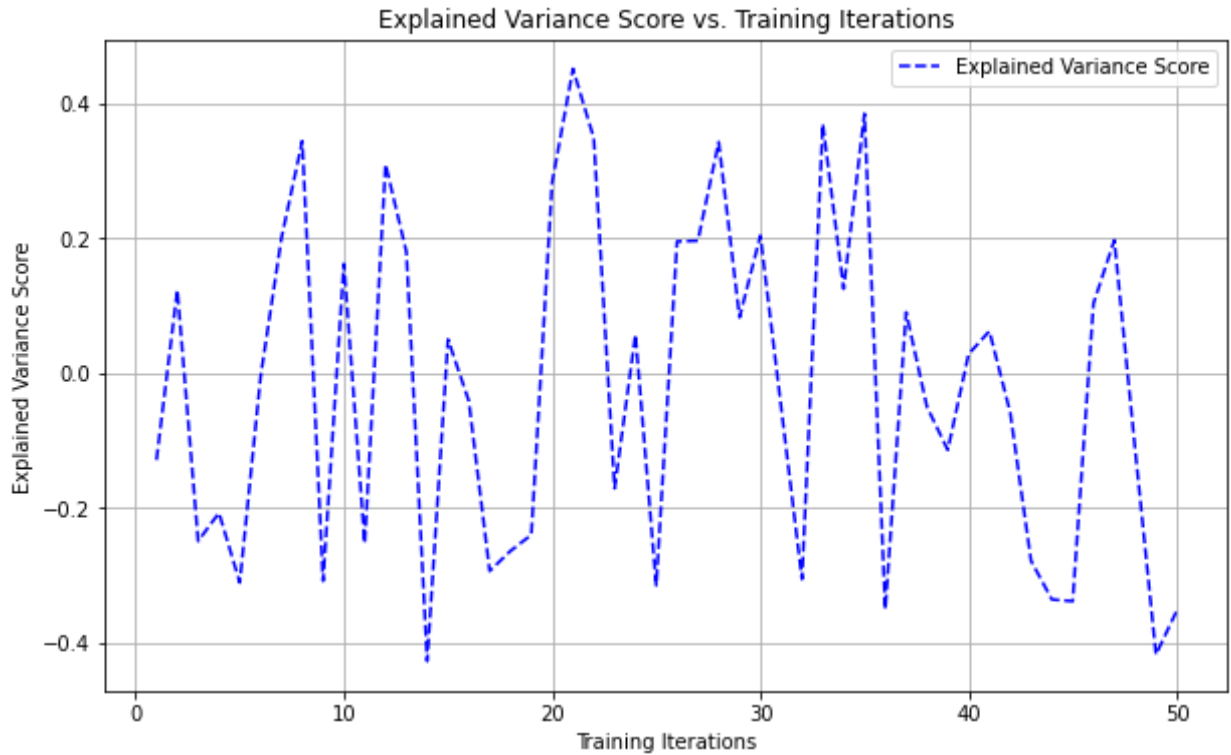

Figure 8: Explained variance score vs training iterations of QNN algorithm.

*4.2 Q-Forest Algorithm*

The Q-Forest algorithm, a quantum-inspired approach, is employed for clustering and classification tasks in large-scale data processing. Drawing upon principles of quantum mechanics such as quantum entanglement and quantum superposition, the algorithm enhances both efficiency and effectiveness. It begins by converting classical data into quantum states, assuming a dataset comprising N samples and d features, with each sample represented as a d-dimensional vector. Subsequently, the algorithm transforms classical data points into quantum states, generating a quantum superposition of the dataset, which enables concurrent manipulation of all data points. The algorithm calculates pairwise distances between data points using a distance metric to establish a quantum entanglement representation of the data as shown in Figure 9. These pairwise distances are encoded into quantum states, such that the entangled states depict the relationships among data points. This encoding method permits the algorithm to evaluate distances between data points more effectively than classical approaches.

The core procedure of the Q-Forest algorithm entails the construction of a quantum decision tree, in which data points are recursively divided into subsets until a termination criterion is satisfied. The quantum decision tree is formed by determining the optimal split point for each node to minimize an impurity measure. Quantum parallelism is employed during this process to efficiently explore all potential split points. Upon completion of the tree, it can be utilized to cluster or classify novel data points based on their quantum entanglement representation.

![2D Representation of Data Points and Pairwise Distances]

Figure 9: Calculation pairwise distances between data points using a distance metric to establish a quantum entanglement

The obtained metric features are represented in Table 3.

Table 3: Obtained Metrics features for Q-Forest Algorithm

| MSE | MAE | Explained Variance Score |
| --- | --- | --- |
| 56.905 | 7.479 | 0.2957 |

Figure 10 shows the variation of Explained variance score with the training iterations.

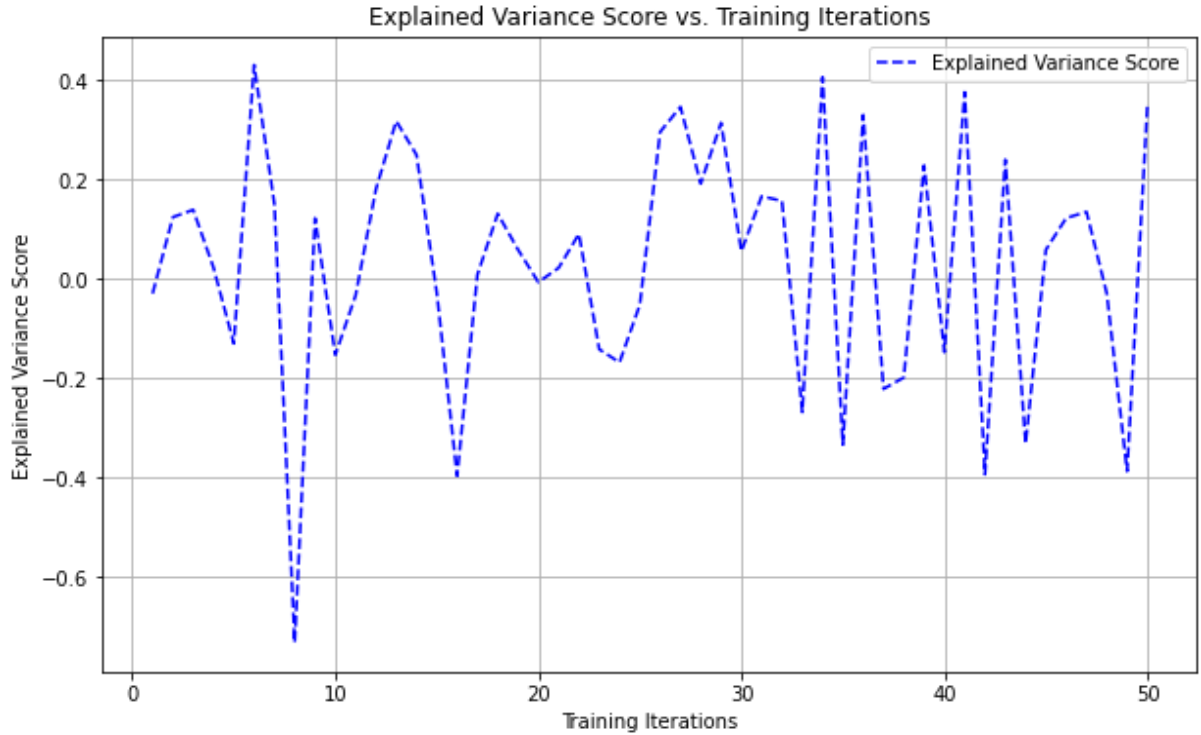

Figure 10: Explained Variance score vs training iterations for Q-Forest algorithm

*4.3 Variational Quantum Classifier (VQC) adapted for regression*

A Variational Quantum Classifier (VQC) adapted for regression is a hybrid quantum-classical machine learning algorithm that uses quantum circuits to perform regression tasks. In this approach, a parametrized quantum circuit is prepared, with adjustable parameters represented by θ. The quantum circuit acts as a feature map, encoding the input data into a high-dimensional quantum state. The encoded data is then processed by a second parametrized quantum circuit, known as the variational circuit, which is responsible for the actual regression. The output of the variational circuit is a continuous value, which is obtained by measuring the expectation value of a specific observable. To adapt VQC for regression, the loss function is tailored to measure the difference between the predicted continuous values and the actual target values. This loss function is used to optimize the parameters of the variational circuit in a classical optimization loop. By iteratively updating the circuit parameters, the VQC model learns the underlying pattern in the data and becomes capable of making predictions for unseen data points.

The obtained metric features are represented in Table 4.

Table 4: Obtained Metrics features for Variational Quantum Classifier (VQC) adapted for regression

| MSE | MAE | Explained Variance Score |
|---|---|---|
| 59.121 | 7.597 | -0.0106 |

Figure 11 shows the variation of Explained variance score with the training iterations.

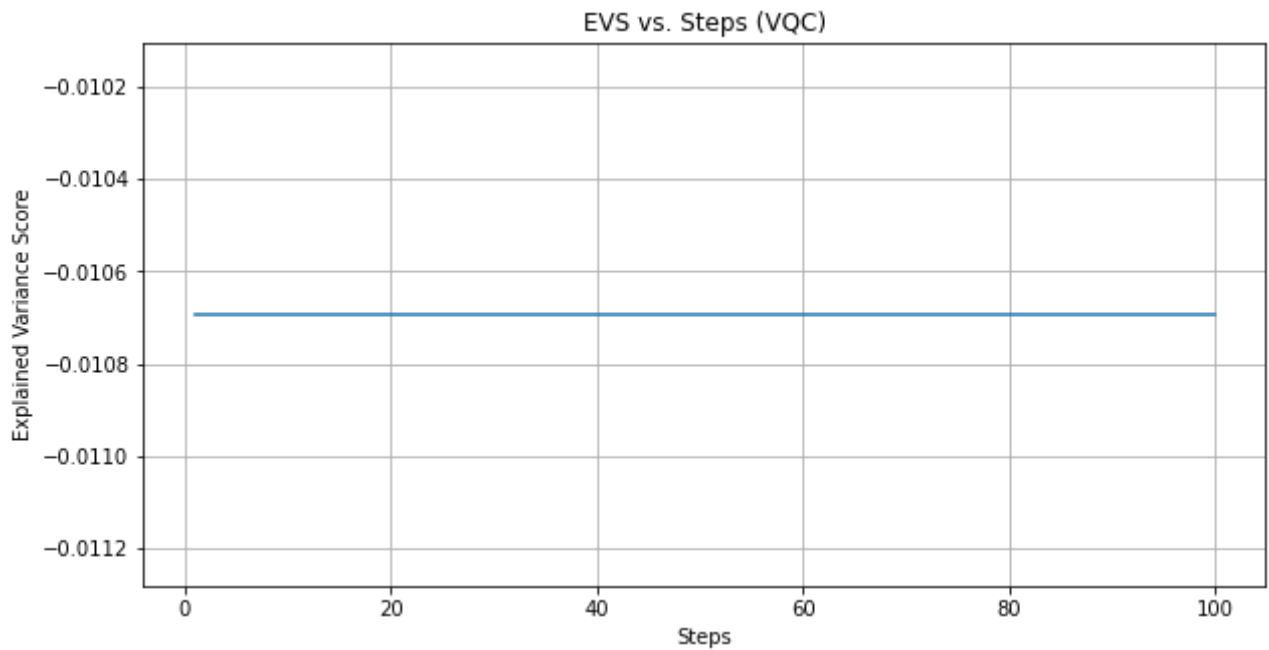

Figure 11: Explained Variance score vs training iterations for Variational Quantum Classifier (VQC) adapted for regression

Figure 12 shows the overall comparison of the obtained results. Our results indicate that the Q-Forest algorithm outperforms the other two algorithms in terms of both MSE and MAE. Q-Forest achieved an MSE of 56.905 and an MAE of 7.479, while the QNN and VQC algorithms recorded an MSE of 60.840 and 59.121, and an MAE of 7.671 and 7.597, respectively. Lower MSE and MAE values indicate better performance in terms of prediction accuracy, demonstrating that Q-Forest is better suited for this particular regression task compared to the other algorithms. Also, the Explained Variance Score (EVS) shows that the Q-Forest algorithm accounts for 29.57% of the total variance in the data, whereas the QNN and VQC algorithms record negative EVS values of -44.4% and -1.06%, respectively. A higher EVS value suggests that the model can better explain the variance in the dataset, and therefore, the Q-Forest algorithm demonstrates superior performance in this aspect as well. These findings suggest that the Q-Forest algorithm is a more effective approach for solving regression tasks compared to the QNN and VQC algorithms, in the context of the dataset and problem studied. However, it is essential to note that the performance of quantum algorithms can vary depending on the specific problem, dataset, and hyperparameter settings. As such, further research is necessary to explore the generalizability of our findings to different datasets and regression tasks. Additionally, it would be interesting to examine the performance of these quantum algorithms in comparison to classical machine learning algorithms, which could offer valuable insights into the advantages and limitations of quantum computing in the field of regression analysis.

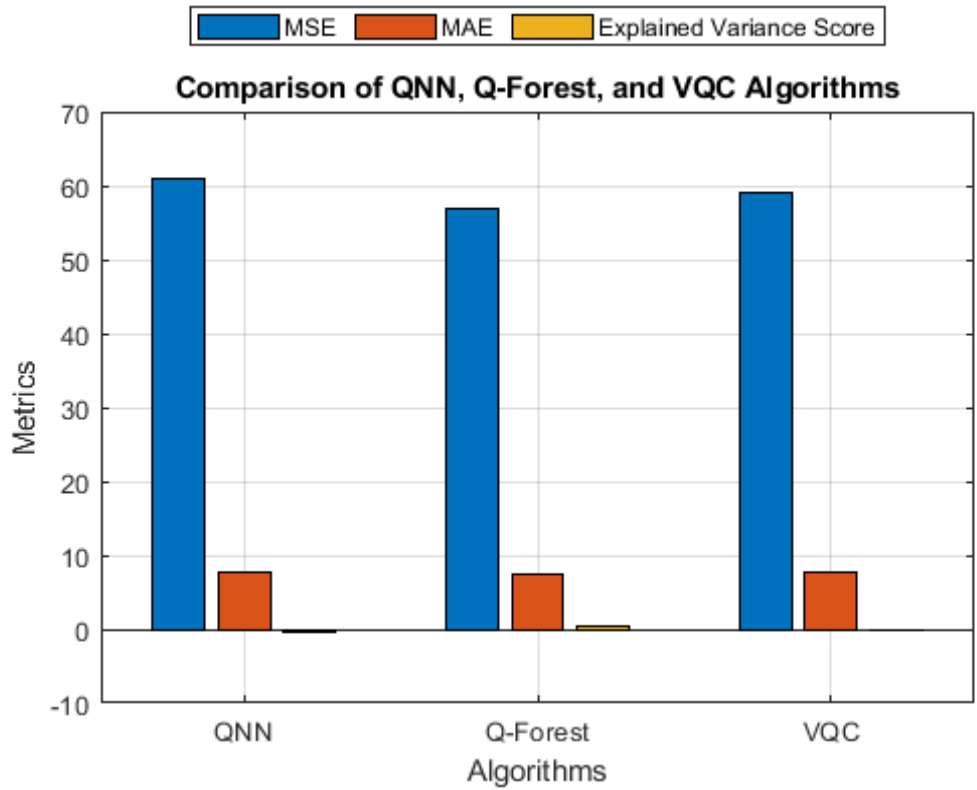

Figure 12: Comparison of the implemented algorithms

## 5. Conclusion

The primary objective of this study was to compare the performance of three quantum algorithms, QNN, Q-Forest, and VQC adapted for regression, for predicting the surface roughness of additive manufactured specimens. Based on our analysis, the Q-Forest algorithm demonstrated the best performance, with an MSE of 56.905, MAE of 7.479, and an EVS of 0.2957. This indicates that the Q-Forest algorithm not only provides more accurate predictions but also accounts for a greater proportion of the variance in the dataset compared to the other two algorithms. The QNN algorithm exhibited a higher MSE of 60.840 and MAE of 7.671, with a negative EVS of -0.444, suggesting that it may not be well-suited for predicting the surface roughness of additive manufactured specimens. Similarly, the VQC adapted for regression achieved an MSE of 59.121, MAE of 7.597, and an EVS of -0.0106, indicating that its performance is also inferior to the Q-Forest algorithm.